\documentclass[preprint,prd,showpacs]{revtex4}
\usepackage{graphicx}
\usepackage{subfigure}

\def\be{\begin{equation}}
\def\ee{\end{equation}}
\def\ben{\begin{eqnarray}}
\def\een{\end{eqnarray}}

\begin{document}

\title{\bf Quantum Lightcone Fluctuations in Compactified Spacetimes}

\author{ Hongwei Yu}
 \email{hwyu@hunnu.edu.cn}
\affiliation{Department of Physics and Institute of  Physics,\\
and Key Laboratory of Low Dimensional Quantum Structure and Quantum Control
of the Ministry of Education, \\
Hunan Normal University, Changsha, Hunan 410081, China }

\author{N.F. Svaiter }
 \email{nfuxsvai@cbpf.br}
 \affiliation{Centro Brasileiro de Pesquisas Fisicas-CBPF \\
 Rua Dr. Xavier Sigaud 150 \\
Rio de Janeiro, RJ, 22290-180, Brazil}

\author{L.H. Ford}
 \email{ford@cosmos.phy.tufts.edu}
\affiliation{Institute of Cosmology, Department of Physics and Astronomy\\
    Tufts University, Medford, MA 02155, USA}



\begin{abstract}
We treat the effects of compactified spatial dimensions on the
propagation of light in the uncompactified directions in the context
of linearized quantum gravity. We find that the flight times of
pulses can fluctuate due to modification of the graviton vacuum by
the compactification. In the case of a five dimensional Kaluza-Klein
theory, the mean variation in flight time can grow logarithmically
with the flight distance. This effect is in principle observable,
but too small to serve as a realistic probe of the existence of
extra dimensions. We also examine the effect of the compactification
on the widths of spectral lines, and find that there is a small line
narrowing effect. This effect is also small for compactification
well above the Planck scale, but might serve as a test of the
existence of extra dimensions.
\end{abstract}


\maketitle 
\baselineskip=14pt

\section{ Introduction}

One of the key features expected in quantum theories of gravity is
fluctuation of the classical lightcone and possible effects on
signal propagation. This possibility has been discussed in several
contexts by numerous
authors~\cite{Pauli,Deser,DeWitt,HS98,NV00,AC00,EMN00,BF04a,BF04b,PR05,CNV06}.
One approach is to study variations in the flight times of light
pulses between a source and a
detector~\cite{Ford95,FS96,YUF,YF00,YW03}. This approach was used in
Ref.~\cite{YUF} to study the effects of boundaries and periodic
compactification of one space dimension. In particular, it was found
that the fluctuation in flight time of a pulse propagating parallel
to a plane boundary, or in a direction transverse to the
compactified dimension, will tend to grow as the flight distance
increases. This growth can be viewed as a cumulative effect of
spacetime geometry fluctuations modified by the boundary or the
compactification.

This effect is analogous to the local Casimir effect, whereby a
boundary modifies the two-point function of the quantized
electromagnetic field. This modification can produce observable
effects, such as a force on an atom. This force was predicted
theoretically by Casimir and Polder~\cite{CP} in 1948, and measured
experimentally by Sukenik~{\it et al}~\cite{Sukenik} in 1993.
Another effect of modified electromagnetic vacuum fluctuations can
be Brownian motion of charged test
particles~\cite{WKF01,YF04,SW08,BBF08}. The presence of a boundary
can alter the mean squared velocity or position of the particle.
This modification is analogous to the effects of lightcone
fluctuations when spacetime geometry fluctuations are modified.

In the present paper, we will be concerned with modification of
lightcone fluctuations due to compact extra dimensions. Theories
with extra dimensions were introduced into physics by
Kaluza~\cite{Kaluza} and Klein~\cite{Klein}, and have been the topic
of many papers in recent years~\cite{PL05}. We wish to address the
question of whether compact extra dimensions can give rise to
observable effects on the propagation of light rays in the
uncompactified directions. Such an effect could be a potential test
for the existence of extra dimensions. A secondary purpose of this
paper is to discuss quantization of linearized gravity in arbitrary
numbers of flat spacetime dimensions and to give explicit expression
for the graviton two-point functions in the transverse tracefree
gauge. In effect, we are searching for modifications of quantum
effects in the uncompactified dimensions due to the presence of the
extra dimensions. A somewhat different effect, whereby extra
dimensions lead to changes in Casimir forces has recently been
discussed by Cheng and others~\cite{Cheng,FK09}.

 The outline of this paper is as follows: In Sect.~\ref{sec:lc_flucts}, we review how quantized
linear metric perturbations may give rise to variations in flight
times of pulses.
 In Sect.~\ref{sec:two-point}, we discuss quantization of linearized metric
perturbations and compute the graviton two-point function in the
transverse, tracefree gauge for flat spacetime of arbitrary
dimension.  In particular, in Sect.~\ref{sec:del_t} we give a new
derivation of the formula for $\Delta t$, the mean flight time
variation.
 This result is used in Sect.~\ref{sec:5DKK} to study
lightcone fluctuations in  a five dimensional spacetime with one
compact dimension. Some results for more than one compact dimension
are also summarized. We also examine the issue of the correlation of
successive pulses in this model in Sect.~\ref{sec:corr}. We turn to
a different measure of lightcone fluctuations, the possible
broadening of spectral lines, in Sect.~\ref{sec:red}. Our results
are discussed in Sect.~\ref{sec:diss}. Appendix~\ref{sec:2pt}
provides a detailed treatment of the graviton two-point functions.


\section{Light Cone Fluctuations and Flight Time Variations }
\label{sec:lc_flucts}


  To begin, let us examine a $d=4+n$ dimensional flat spacetime with $n$  extra dimensions.
 Consider a flat background  spacetime  with a linearized perturbation
$h_{\mu\nu}$ propagating upon it , so the spacetime metric may be
written as
\begin{equation}
ds^2  = (\eta_{\mu\nu} +h_{\mu\nu})dx^\mu dx^\nu = dt^2 -d{\bf x}^2
+ h_{\mu\nu}dx^\mu dx^\nu \, ,
\end{equation}
where the indices $\mu,\nu$ run through $0,1,2,3,...,3+n$. Let
$\sigma(x,x')$ be one half of the squared geodesic distance between
 a pair of spacetime points $x$ and $x'$,   and $\sigma_0(x,x')$
be the corresponding quantity in the flat background . In the
presence of a linearized metric  perturbation,
 $h_{\mu\nu}$, we may expand $
\sigma = \sigma_0 + \sigma_1 + O(h^2_{\mu\nu}) \, .$ Here $\sigma_1$
is first order  in $h_{\mu\nu}$. If we quantize  $h_{\mu\nu}$, then
quantum gravitational vacuum fluctuations will  lead to fluctuations
in the geodesic separation, and therefore induce
 lightcone fluctuations.  In particular, we have $\langle \sigma_1^2 \rangle \not= 0$,
since $\sigma_1$  becomes a quantum operator when the metric
perturbations are quantized. The quantum lightcone fluctuations give
rise to fluctuations in the speed of light, which may produce a time
delay or advance $\Delta t$ in the arrival times of pulses.

We are concerned with how lightcone fluctuations characterized by
 $\langle \sigma_1^2 \rangle$ are related to physical observable quantities. For
this purpose, let us  consider the propagation of light pulses
between a source and a detector separated by a distance $r$ on a
flat background with quantized linear perturbations. For a pulse
which is delayed or advanced by time $\Delta t$, which is much less
than $r$, one finds \be
\sigma=\sigma_0+\sigma_1+....={1\over2}[(r+\Delta t)^2-r^2]\approx
r\Delta t\,. \label{eq:separation} \ee Square the above equation and
take the average over a given  quantum state of gravitons
$|\phi\rangle$ (e.g. the vacuum states associated with
compactification of spatial dimensions ), \be \Delta
t_{\phi}^2={\langle \phi| \sigma_1^2 |\phi\rangle\over r^2}\,.
\label{eq:TPHI} \ee This result is, however,  divergent due to the
formal divergence of
 $\langle\phi| \sigma_1^2 |\phi\rangle $. One can
 define
 an observable $\Delta t$  by subtracting from Eq.~(\ref{eq:TPHI}) the corresponding
quantity, $\Delta t_0^2$, for the  vacuum state as follows \be
 \Delta t^2=\Delta t_{\phi}^2-\Delta t_0^2 =
{\langle \phi| \sigma_1^2 |\phi\rangle-\langle 0| \sigma_1^2
|0\rangle \over r^2}\equiv {\langle \sigma_1^2 \rangle_R\over r}\,.
\ee In this case, we are dealing with the shift in the light cone
fluctuations due to a change in quantum state or spacetime topology.
We do not attempt to treat the vacuum state of uncompactified
Minkowski spacetime, but rather the dependence of the light cone
fluctuations on some parameter which can be varied.
 Therefore, the root-mean-squared  deviation from the classical
propagation time is given by
 \be \Delta t= {\sqrt{\langle
\sigma_1^2 \rangle_R}\over r}\,. \label{eq:MDT} \ee
 Note that
$\Delta t$ is the ensemble averaged deviation, not necessarily the
expected variation in flight time, $\delta t $,  of two pulses
emitted close together in time. The latter is given by $\Delta t$
only when the correlation time between successive pulses is less
than the time separation of the pulses. This can be understood
physically as due to the fact that the gravitational field may not
fluctuate significantly in the
 interval between the two pulses.
This point is discussed in detail in Ref.~\cite{FS96}. These
stochastic fluctuations in the apparent velocity of light arising
from
 quantum gravitational
fluctuations are in principle observable, since they may lead to a
spread in the arrival times of pulses from distant sources.

  In order to find $\Delta t$ in a particular situation,
we need to calculate the quantum expectation value
 $\langle \sigma_1^2 \rangle_R$ in any
chosen quantum state $|\psi\rangle$, which can be shown to be given
by~\cite{Ford95,YUF} \be \langle \sigma_1^2 \rangle_R ={1\over
8}(\Delta r)^2 \int_{r_0}^{r_1} dr \int_{r_0}^{r_1} dr' \:\,
n^{\mu} n^{\nu} n^{\rho} n^{\sigma} \:\,
G^{R}_{\mu\nu\rho\sigma}(x,x') \,. \label{eq:interval} \ee Although
the previous derivations in Ref.~\cite{YUF}  were given in 3+1
dimensions, the generalization to arbitrary dimensions is
straightforward. Here $ dr=|d{\bf x}|$,  $\Delta r=r_1-r_0$ and $
n^{\mu} =dx^{\mu}/dr$. The integration is taken along the null
geodesic connecting two points $x$ and $x'$, and \be
 G^{R}_{\mu\nu\rho\sigma}(x,x')= \langle \psi| h_{\mu\nu}(x) h_{\rho\sigma}(x')+
 h_{\mu\nu}(x') h_{\rho\sigma}(x)|\psi \rangle
\ee is the graviton Hadamard  function, understood to be suitably
renormalized. The gauge invariance of  $\Delta t$, as given by
Eq.~(\ref{eq:MDT}), was analyzed in Ref.~\cite{YUF}. An alternative
derivation which makes the gauge invariance more obvious is given in
Sect.~\ref{sec:del_t}.


\section{Quantization in the Transverse Tracefree Gauge }
\label{sec:two-point}


\subsection{Minkowski Spacetimes}

We will use a quantization of the linearized gravitational
perturbations $h_{\mu\nu}$ in flat spacetime with arbitrary
dimension which retains only physical degrees of freedom. That is,
we are going to work in a transverse, tracefree (TT) gauge defined
by
\begin{equation}
h = h^\mu_\mu =0\,, \quad \partial_\mu\, h_{\mu\nu} =0\,, \quad {\rm
and}\; u^\mu \, h_{\mu\nu} =0 \,,
\end{equation}
where $u^\mu$ is a time-like vector. In the frame of reference in
which $u^\mu=(1,0,0,0)$, the gravitational perturbations have only
spatial components $h_{ij}$,
 satisfying the transverse,  $\partial^i h_{ij}=0$,  and tracefree, $ h^i_i=0$
conditions. Here $i,j$ run from 1 to $3+n=d-1$.  These $2d$
conditions remove all of the gauge degrees of freedom and leave
${1\over 2}(d^2-3d)$ physical degrees of freedom. We write the
quantized gravitational perturbation operator as
\begin{equation}
h_{ij} = \sum_{{\bf k},\lambda}\, [a_{{\bf k}, \lambda} e_{ij}
({{\bf k}, \lambda})
  f_{\bf k} + H.c. ].
\end{equation}
Here H.c. denotes the Hermitian conjugate, $\lambda$ labels the
${1\over 2}(d^2-3d)$ independent
 polarization states,   $f_{\bf k}$  is the mode function,
 and the $e_{\mu\nu} ({{\bf k}, \lambda})$ are polarization tensors. The graviton
creation and annihilation operators satisfy the usual commutation
relation:
\begin{equation}
[a_{{\bf k}, \lambda},a^\dagger_{{\bf k'}, \lambda'}] = \delta_{{\bf
k},{\bf k'}}\, \delta_{\lambda,\lambda'}\,.
\end{equation}
This relation may be taken to be the fundamental quantization
postulate. Units in which $32\pi G_d =1$, where $G_d$ is Newton's
constant in $d$ dimensions,  and in which $\hbar =c =1$  will be
used in this paper, except as otherwise noted.

Let us now calculate the  Hadamard function,
$G_{\mu\nu\rho\sigma}(x,x')$,
 for gravitons in the Minkowski vacuum state in the transverse tracefree gauge.
(By Minkowski we mean flat spacetime with all dimensions
uncompactified.) It follows that
\begin{equation}
 G_{ijkl}(x,x')=\frac{2 Re}{(2\pi)^{d-1}}\int\,{d^{d-1}{\bf k}
\over{2 \omega}} \sum_{\lambda} \, e_{ij} ({{\bf k}, \lambda})
e_{kl} ({{\bf k}, \lambda}) e^{i{\bf k} \cdot({\bf x}-{\bf
x'})}e^{-i\omega(t-t')}\,.
\end{equation}
The summation of polarization tensors in the transverse tracefree
gauge can be found  using the tensorial argument in the Appendix of
Ref.~\cite{YUF}.
\begin{eqnarray}
\sum_{\lambda}\,&& e_{ij} ({{\bf k}, \lambda})\, e_{kl} ({{\bf k},
\lambda})= \delta_{ik}\delta_{jl} +\delta_{il}\delta_{jk}-{2\over
d-2} \delta_{ij}\delta_{kl}
+{2(d-3)\over d-2}\hat k_i\hat k_j \hat k_k\hat k_l\nonumber\\
&& +{2\over d-2}\biggl(\hat k_i \hat k_j \delta_{kl} +\hat k_k \hat
k_l \delta_{ij}\biggr)-\hat k_i \hat k_l \delta_{jk} -\hat k_i \hat
k_k \delta_{jl}-\hat k_j \hat k_l \delta_{ik}-\hat k_j \hat k_k
\delta_{il}\,,
\end{eqnarray}
where $\hat k_i=\frac{ k_i}{ k}\,.$  We find that
\begin{eqnarray}
 G_{ijkl}&=& {2\over d-2}\biggl(2F_{ij}\delta_{kl}
+2F_{kl}\delta_{ij}\biggr)-2F_{ik}\delta_{jl}-2F_{il}
\delta_{jk}-2F_{jl}\delta_{ik} -2F_{jk}\delta_{il}\nonumber\\
&&+{4(d-3)\over d-2}H_{ijkl}
 +2D(x,x')\biggl(\delta_{ik}\delta_{jl}+\delta_{il}\delta_{jk}-{2\over
d-2}\delta_{ij} \delta_{kl}\biggr)\,. \label{eq:G}
\end{eqnarray}
Here $D(x,x')$,  $F_{ij}(x,x')$ and $H_{ijkl}(x,x')$ are functions
which are
 defined as follows:
\be
    D^n(x,x')=
{{\rm Re}\over{(2\pi)^{3+n}}}\int\, {d^{3+n}{\bf k}\over{2
\omega}}e^{i{\bf k} \cdot({\bf x}-{\bf x'})}e^{-i\omega(t-t')} \,,
\label{eq:Dfunc} \ee

\be
    F^n_{ij}(x,x')={{\rm Re}\over{(2\pi)^{3+n}}}\partial_i\partial'_j\int\,
{d^{3+n}{\bf k}\over{2 \omega^3}}e^{i{\bf k} \cdot({\bf x}-{\bf
x'})} e^{-i\omega(t-t')} \,, \label{eq:Ffunc} \ee and \be
    H^n_{ijkl}(x,x')={{\rm Re}\over{(2\pi)^{3+n}}}
   \partial_i\partial^{\prime}_j\partial_k\partial_l^{\prime}
\int\, {d^{3+n}{\bf k}\over{2 \omega^5}}e^{i{\bf k} \cdot({\bf x}-
{\bf x'})}e^{-i\omega(t-t')} \,. \label{eq:Hfunc} \ee These
functions are calculated in Appendix~\ref{sec:2pt}.


\subsection{Flat Spacetimes with Periodic Compactification}
\label{sec:periodic}


Let us now suppose that  the extra $n$ dimensions $z_1,...,z_n$ are
compactified with
 periodicity lengths $L_1,...,L_n$, namely spatial points $z_i$ and $z_i+L_i$ are
identified.  For simplicity, we shall assume in this paper that
$L_1=...=L_n=L$. The effect of imposition of the periodic boundary
conditions on the extra dimensions is to restrict the field modes to
a discrete set \be f_{\bf k} = (2\omega (2\pi)^{3}L^n)^{-{1\over 2}}
e^{i({\bf k \cdot x} -\omega t)} \, ,
    \label{eq:mode2}
\ee with \be k_{i}=\frac{2\pi m_i}{L}, \quad i=1,...,n,\quad
m_i=0,\pm 1, \pm 2, \pm 3,...\,. \ee Let us denote the associated
vacuum state by $|0_L\rangle$.
 In order to calculate the
gravitational vacuum fluctuations due to compactification of  extra
dimensions, we need the renormalized graviton Hadamard function with
respect to the vacuum state $|0_L\rangle$,
$G^{R}_{\mu\nu\rho\sigma}(x,x')$, which is given by a multiple
 image sum of the
corresponding Hadamard function for the Minkowski  vacuum,
$G_{\mu\nu\rho\sigma}$ : \be
 G_{\mu\nu\rho\sigma}^{ R}(t, z_i,\,t',z_i')
 =\prod_{i=1}^n{\sum_{m_i=-\infty}^{+\infty}}^{\prime}
G_{\mu\nu\rho\sigma}(t, z_i   , \,t',z_i' +m_iL)\, . \ee Here the
prime on the summation indicates that the $m_i=0$ term is excluded
and the notation \be (t,\vec{x}, z_1,.., z_{n})\equiv(t, z_{i}) \ee
 has been adopted.

We are mainly concerned about how lightcone fluctuations arise in
the usual uncompactified space as a result of compactification of
extra dimensions. So we shall examine the case of a light ray
propagating in one of the uncompactified dimensions. Take the
direction to be along the $x$-axis in our four dimensional world,
then the relevant graviton two-point function is $G_{xxxx}$, which
can be expressed as \ben G_{xxxx}(t,\vec{x},z_i,\,t',\vec{x}',z_i')
&=& {4(n+1)\over n+2} \biggl[ D(t,\vec{x},z_i,\,t',\vec{x}',z_i')
 - 2 F_{xx} (t,\vec{x},z_i,\,t',\vec{x}',z_i')\nonumber\\
&& + H_{xxxx}(t,\vec{x},z_i,\,t',\vec{x}',z_i') \biggr]\,.
\label{eq:Gx} \een Assuming that the propagation goes from point
$(a,0,...,0)$ to point $(b,0,...,0)$, we have \ben \langle
\sigma_1^2 \rangle &=&{1\over 8}(b-a)^2 \int_{a}^{b} dx \int_{a}^{b}
dx'\, G_{xxxx}^{R }(t,x,{\bf 0},\,t',x',{\bf 0},),
 \,\nonumber\\
&=& {1\over 8}(b-a)^2 \int_{a}^{b} dx \int_{a}^{b} dx'\,
\prod_{i=1}^n{\sum_{m_i=-\infty}^{+\infty}}^{\prime}
G_{xxxx}(t,x,{\bf 0},\,t',x',0,0, m_1L,...,m_iL)\,.\nonumber\\
\label{eq:sigma1} \een With  these results, we can in principle
calculate lightcone fluctuations in spacetimes with an arbitrary
number of flat extra dimensions. Recall that we are working in units
in which the Newton's constant in $d$-dimensions is $G_d = (32
\pi)^{-1}$. In final results, it will be useful to convert to more
familiar units using the relation that \be \ell_P^2 = G_4 = G_d\,
L^{4-d} \, ,   \label{eq:Newton} \ee where $\ell_P \approx 10^{-34}
{\rm cm}$ is the Planck length.


\subsection{An Alternative Derivation of $\Delta t$}
\label{sec:del_t}

In this section, we wish to rederive $\Delta t$ using the geodesic
deviation equation. This derivation allows us to see the gauge
invariance more clearly, and to discuss the issue of Lorentz
invariance of lightcone fluctuations.
 Let us consider a pair of  timelike geodesics with tangent vector $u^{\mu}$,
and $n^{\mu}$ as a unit spacelike vector pointing from one geodesic
to the other (See Fig. 1).
\begin{figure}[hbtp]
\begin{center}
\includegraphics[height=6cm]{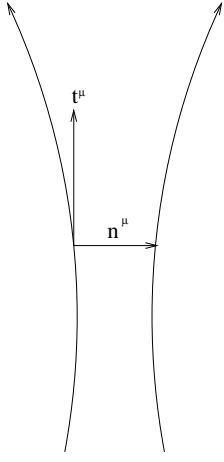}
\end{center}
\caption{ A pair of nearby timelike geodesics. Here  $u^{\mu}$ is a
tangent vector along the geodesic, while $n^{\mu}$ is  a unit
spacelike vector pointing from one geodesic to the other.  }
\label{geodesic}
\end{figure}
The geodesic deviation equation is given by \be {D^2n^{\mu}\over
d\tau^2} = -R^{\mu}_{\alpha\nu\beta}u^{\alpha}n^{\nu}u^{\beta}\,,
\ee where $R^{\mu}_{\alpha\nu\beta}$ is the Riemann tensor. The
relative acceleration per unit proper length of particles on the
neighboring geodesics is \be \alpha \equiv n_{\mu}{D^2n^{\mu}\over
d\tau^2}= -R_{\mu
\alpha\nu\beta}n^{\mu}u^{\alpha}n^{\nu}u^{\beta}\,. \ee Thus if $ds$
is the spatial distance between the two particles, then $\alpha\,ds$
is their relative acceleration. It follows that the relative change
in displacement of the two particles after a proper time $T$ is \be
ds\,\int^T_0\,d\tau\,\int^{\tau}_0\,d\tau' \alpha(\tau',0) \,, \ee
Now consider the case of two observers (particles) separated by a
finite initial distance $s_0$ as illustrated in
Fig.~\ref{geodesic2}.

\begin{figure}[hbtp]
\begin{center}
\includegraphics[width=6cm]{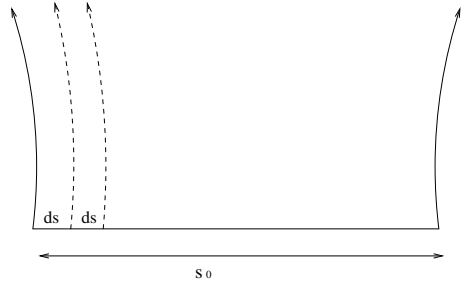}
\end{center}
\caption{ Two timelike geodesics separated by a finite interval
containing an infinite number of nearby geodesics .}
\label{geodesic2}
\end{figure}
 We can find the relative change in displacement of these two observers by
integrating on $s$: \be \Delta s =
\int^{s_0}_0\,ds\,\int^T_0\,d\tau\,\int^{\tau}_0\,d\tau'
\alpha(\tau',0) \,. \ee This is the relative displacement measured
at the same moment of proper time for both observers.

\begin{figure}[hbtp]
\begin{center}
\includegraphics[height=6cm]{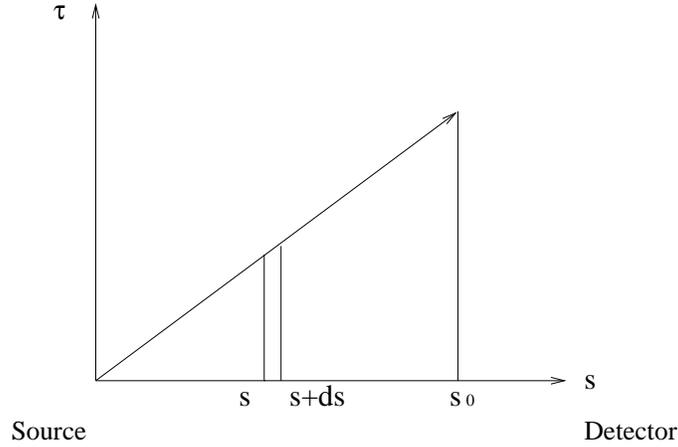}
\end{center}
\caption{The displacement, $\Delta s$, between a source and a
detector is given by an integration within the triangular region. }
\label{fig=geodesic3}
\end{figure}

Let us now consider a light signal sent from one observer to the
other. If $\alpha=0$, the distance traveled by the light ray is
$s_0$. When $\alpha\neq 0$, this distance becomes $s_0+\Delta s$,
where now \be \Delta s =
\int^{s_0}_0\,ds\,\underbrace{\int^s_0\,d\tau\,\int^{\tau}_0\,
d\tau' \alpha(\tau',s)} \,. \ee Here the  under-braced integral is
the displacement per unit $s$ of a pair of
 observers at a distance $s$ from the source. The domain of the final two
integrations is illustrated in Fig.~\ref{fig=geodesic3}.

If gravity is quantized, the Riemann tensor will fluctuate around an
average value of zero due to quantum gravitational vacuum
fluctuations. This leads to $\langle \alpha \rangle=0 $, and hence
 $\langle \Delta s \rangle=0$. Notice here
that   $ \alpha$ becomes a quantum operator when metric
perturbations are
 quantized.
However, in general,  $\langle (\Delta s )^2\rangle\neq0$,  and we
have \be
 \langle (\Delta s )^2\rangle=
\int^{s_0}_0\,ds_1\int^{s_0}_0\,ds_2\int^{s_1}_0\,d\tau_1\int^{\tau_1}_0\,
d\tau'_1\int^{s_2}_0\,d\tau_2\int^{\tau_2}_0 \,d\tau'_2\,\langle
\alpha(\tau'_1,s_1)\alpha(\tau'_2,s_2)\rangle\,. \label{eq:MDD} \ee
Thus the root-mean-squared fluctuation in the flight path is
$\sqrt{\langle (\Delta s )^2\rangle}$, which can also be understood
as a fluctuation in the speed of light. It entails an intrinsic
quantum uncertainty in the measurement of distance. Therefore,
spacetime becomes fuzzy at a scale characterized by $\sqrt{\langle
(\Delta s )^2\rangle}$.
 The integrand in Eq.~(\ref{eq:MDD}) is obviously invariant under any
 coordinate transformation while the
integral is gauge invariant within the linear approximation.

We now wish to show that this gauge-invariant quantity is the same
as Eq.~(\ref{eq:MDT}) when calculated in the transverse-tracefree
(TT) gauge. Choose a coordinate system where the source and the
detector are both at rest, and suppose that the light ray propagates
in the $x$-direction, then we have \be u^{\mu}=(1,0,0,0)\,, \ee \be
n^{\mu}=(0,1,0,0)\,, \ee and \be \alpha
=R_{xtxt}=-{1\over2}h_{xx,tt}\,.  \label{eq:Rxtxt} \ee Substitution
of the above results into Eq.~(\ref{eq:MDD}) leads to
 \ben
 \langle (\Delta s )^2\rangle &=&
\int^{r}\,dx_1\int^{r}_0\,dx_2\int^{x_1}_0\,dt_1\int^{t_1}_0\,dt'_1
\int^{x_2}_0\,dt_2\int^{t_2}_0
\,dt'_2\,\langle \alpha(t'_1,s_1)\alpha(t'_2,s_2)\rangle \nonumber\\
&=&{1\over 4}\,\int^r_0\,dx_1\int^r_0\,dx_2\, \langle
h_{xx}(x_1,x_1) h_{xx}(x_2,x_2)\rangle={1\over r} \langle
\sigma_1^2\rangle\,, \een where we have set $s_0=r$ and used the
fact that along the light ray $x=t$. Thus, one has \be \Delta t= {
\sqrt{\langle \sigma_1^2\rangle}\over r}=\sqrt{\langle (\Delta s
)^2\rangle } \ee which also demonstrates the gauge-invariance of
$\Delta t$.

Now we wish to discuss the rather subtle issue of the relation of
lightcone fluctuations to Lorentz symmetry. It is sometimes argued
that lightcone fluctuations are incompatible with Lorentz
invariance. The most dramatic illustration of this arises when a
time advance occurs, that is, when a pulse propagates outside of the
classical lightcone. In a Lorentz invariant theory, there will exist
a frame of reference in which the causal order of emission and
detection is inverted, so the pulse is seen to be detected before it
was emitted. Thus the lightcone fluctuation phenomenon, if it is to
exist at all, seems to be incompatible with strict Lorentz
invariance.

Our view of the situation is the following: lightcone fluctuations
respect Lorentz symmetry on the average, but not in individual
measurements. The symmetry on the average insures that the mean
lightcone be that of classical Minkowski spacetime. The average
metric is that of Minkowski spacetime provided that $\langle
h_{\mu\nu} \rangle = 0.$ However, a particular pulse effectively
measures a spacetime geometry which is not  Minkowskian and not
Lorentz invariant. A simple model may help to illustrate this point.
Consider a quantum geometry consisting of an ensemble of classical
Schwarzschild spacetimes, but with both positive and negative values
for the mass parameter $M$. (The fact that the $M < 0$ Schwarzschild
spacetime has a naked singularity at $r=0$ need not concern us. For
the purpose of this model, we can confine our discussion to a region
where $r \gg |M|.$) Suppose that this ensemble has $\langle M
\rangle= 0$, but $\langle M^2 \rangle \not= 0$. It is well known
that light propagation in a $M > 0$ Schwarzschild spacetime can
exhibit a time delay relative to what would be expected in flat
spacetime. This is the basis for the time delay tests of general
relativity using radar signals sent near the limb of the sun. In the
present model, however, the time difference is equally likely to be
a time advance rather than a time delay. A measurement of the time
difference amounts to a measurement of $M$. This model is Lorentz
invariant on the average because $\langle M \rangle= 0$ and the
average spacetime is Minkowskian. However, a specific measurement
selects a particular member of the ensemble, which is generally not
Lorentz invariant.

In addition to the fact that the mean metric is Minkowskian, there
is another sense in which lightcone fluctuations due to
compactification exhibit average Lorentz invariance. Note  that
$\Delta s$, and hence $\Delta t$, depends on the Riemann tensor
correlation function $\langle R_{xtxt}(x_1)R_{xtxt}(x_2)\rangle$,
which is invariant under Lorentz boosts along the $x$-axis. Thus if
we were to repeat
 the above
calculations of $\Delta s$ in a second frame moving with respect to
the first, the result will be the same. In both cases one is
assuming that the detector is at rest relative to the source. This
is a reflection of the Lorentz invariance of the spectrum of
fluctuations, which is exhibited by the compactified flat spacetimes
studied in this paper, but not by the
 Schwarzschild spacetime with a fluctuating mass.


\section{ The five dimensional Kaluza-Klein model}
\label{sec:5DKK}


In this section, we will specialize to the case of one extra
compactified dimension, so $d=5$ and $n=1$. This corresponds to the
original Kaluza-Klein model~\cite{Kaluza,Klein}.

\subsection{Calculation of $\Delta t$}

To begin, let us examine the influence of the compactification of
the fifth (extra) dimension on the light propagation in our four
dimensional world, by considering a light ray traveling along the
$x$-direction from point $a$ to point $b$, which is perpendicular to
the direction of compactification. Define \be \rho=x-x',\quad\quad
b-a=r\, \ee and note the fact that the integration in
Eq.~(\ref{eq:sigma1}) is to be carried out along the classical null
geodesic on which $ t-t' =\rho$. To calculate $\Delta t$, we need
the graviton two-point function component $G_{xxxx}$, which in this
case is given by Eq.~(\ref{eq:G}) as \be G_{xxxx} = \frac{8}{3}\,(D
-2 F_{xx} + H_{xxxx})\,. \ee The quantities in this espression may
be computed from Eqs.~(\ref{eq:Dn}), (\ref{eq:F}), (\ref{eq:S0}),
(\ref{eq:H}), and (\ref{eq:Q0}), with the result \be
G_{xxxx}(t,x,0,0,0, t',x',0,0,mL')= {8\over 3\pi^2}{\rho^4 m L (5
m^2 L^2-3\rho^2)\over(\rho^2+ m^2L ^2)^5} \,. \ee

Thus, we have
 \ben
 \langle \sigma_1^2 \rangle_R &=&{1\over
8}r^2 \int_{a}^{b} dx \int_{a}^{b} dx'\, G_{xxxx}^{R }(t,x,{\bf
0},\,t',x',{\bf 0},),
 \,\nonumber\\
&=& {1\over 8}r^2 \int_{a}^{b} dx \int_{a}^{b}
dx'\,{\sum_{m=-\infty}^{+\infty}}^{\prime}
G_{xxxx}(t,x,{\bf 0},\,t',x',0,0, mL)\nonumber\\
& =&{2 r^2\over 9\pi^2 L}{\sum_{m=1}^{\infty}}\, {\gamma^6 \over m
(m^2+\gamma^2)^3}\,,
 \een
where we have introduced a dimensionless parameter $\gamma=r/L$. We
are interested here in the case in which $ \gamma \gg 1$. Thus the
summation can be approximated by integration as follows
 \be
\langle \sigma_1^2 \rangle_R={2 r^2\over 9\pi^2
L}\,\int_{1/\gamma}^{\infty}\,{1\over x (x^2+1)^3}\,dx
 \ee
which leads to
 \be \langle
\sigma_1^2 \rangle_R \approx {2r^2\over 9\pi^2L} \ln { r \over L}\,.
\label{eq:sigmaFix} \ee

Thus the mean deviation from the classical propagation time due to
the lightcone fluctuations is \be \Delta t \approx \sqrt{ {2\over
9\pi^2L} \ln { r \over L}}=\sqrt{ {2\over 9\pi^2L} \ln { r \over
L}}\sqrt{32\pi G_5} =\sqrt{ {64\over9\pi}}\, \ell_P\, \sqrt{\ln { r
\over L}}\,, \label{eq:tkk} \ee where we have used
Eq.~(\ref{eq:Newton}). This result reveals that the mean deviation
in the arrival time increases logarithmically  with $r$, which
contrasts with the square root growth in the four dimensional case
with one compactified spatial dimension~\cite{YUF}. It also grows as
the size of the compactified dimension decreases. However, even if
$r$ is of cosmological size and $L$ is near the Planck scale,
$\Delta t$ is never more than a couple of orders of magnitude larger
than the Planck scale and hence unobservable in practice. Note that
Eq.~(\ref{eq:tkk}) corrects an erroneous result in Ref.~\cite{YF00},
where a linear growth of $\Delta t$ was found.

We have used the results in Appendix~\ref{sec:2pt} to perform the
analogous calculation for more than one compact dimension. For up to
seven extra dimensions, the result for $\Delta t$ is of the same
form as Eq.~(\ref{eq:tkk}), a logarithmic growth with $r$.

\subsection{Correlation of Pulses}
\label{sec:corr}

The fluctuation in the flight time of pulses, $\Delta t$, can apply
to the successive pulses. However, $\Delta t$ is the expected
variation in the arrival times of two successive pulses only when
they are uncorrelated~\cite{FS96}.
 To determine the
correlation, we need to  compare $|\langle \sigma_1^2\rangle| $ and
$|\langle \sigma_1\sigma'_1\rangle| $.  The latter quantity is
defined by
\begin{equation}
\langle \sigma_1 \sigma'_1 \rangle = {1\over 8}(\Delta r)^2
\int_{0}^{r} dr_1 \int_{0}^{r} dr_2 \:\,  n^{\mu} n^{\nu} n^{\rho}
n^{\sigma} \:\, G^{R}_{\mu\nu\rho\sigma}(x_1,x_2) \, ,
 \label{eq:sigma11'ex}
\end{equation}
where the $r_1$-integration is taken along the mean path of the
first pulse, and the $r_2$-integration is taken along that of the
second pulse. Here we will assume that $\Delta t \ll  r$, so the
slopes of the two mean paths are approximately unity.
 Let the time separation of the emission of the two pulses be $T$.
Thus the two-point function in Eq.~(\ref{eq:sigma11'ex}) will be
assumed to be evaluated at $\rho = |{\bf x}_1 -{\bf x}_2| = |r_1
-r_2|$ and $\tau = |t_1 - t_2| = |r_1 -r_2 -T|$. If $|\langle
\sigma_1\sigma'_1\rangle| \ll |\langle \sigma_1^2\rangle| $, two
pulses are uncorrelated, and otherwise they are correlated.

In Appendix~\ref{sec:sigma_sigma}, it is shown that \ben \langle
\sigma_1\sigma'_1\rangle \approx  {r^2\over 9\pi^2L} \ln { 2r \over
T}\,. \label{eq:sigma_sigma} \een Compare this result with
 \be \langle
\sigma_1^2 \rangle_R \approx  {2r^2\over 9\pi^2L} \ln { r \over
L}\,. \ee
 We can see that
 two successive pulses separated by $T$ in time are only weakly correlated
($|\langle \sigma_1\sigma'_1\rangle|\ll|\langle \sigma_1^2
\rangle_R|$)
  provided that
   \be r\gg  {2L^2\over T}\,. \ee
equivalently \be T\gg \frac{2L^2}{r}\, . \label{eq:Leff} \ee
However, if $r\ll L$, one can show, by series expansion, that
   $|\langle
\sigma_1\sigma'_1\rangle|\ll|\langle \sigma_1^2 \rangle_R|$, if
$T\gg L$, and $|\langle \sigma_1\sigma'_1\rangle|\approx |\langle
\sigma_1^2 \rangle_R|$, when $T<L$.

A few comments are now in order about the physical picture behind
our correlation results. It is natural to expect from the
configuration that the dominant contributions to the light cone
fluctuation come from the graviton modes with wavelengths of the
order of $\sim L$. In other words, the lightcone fluctuates on a
typical time scale of
 $\sim 1/L$. If the travel
distance, $r$, is less than $L$, successive pulses are uncorrelated
only when their time separation is  greater than the typical
fluctuation time scale. Otherwise they are correlated because the
quantum gravitational vacuum fluctuations are not significant enough
in the interval between the pulses. However, if $r\gg 2L^2/T$, then
successive pulses are in general weakly correlated. Thus the
correlation time for large $r$ is of order $L^2/r$, which is much
smaller than the compactification scale $L$. We can understand this
result as arising from the loss of correlation as the pulses
propagate over an increasing distance.

\section{Redshift Fluctuations}
\label{sec:red}

\begin{figure}
\begin{center}
\includegraphics[width=9cm]{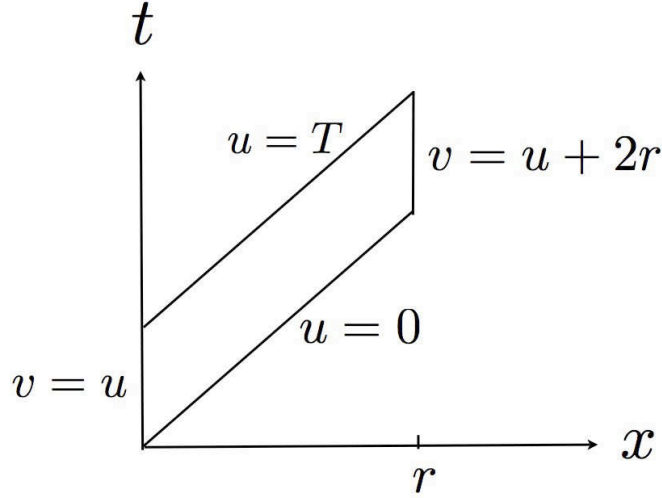}
\end{center}
\caption{The region of integration in the expressions for $\delta
\xi^2$ is illustrated. The worldline of the source is $x=0$, which
is equivalent to $v=u$. Here $u-t-x$ and $v=t+x$. The worldline of
the detector is $x=r$, or $v = u +2r$. The first null ray is emitted
at $t=0$ and travels on the $u=0$ line. the second ray is emitted at
$t=T$ and travels on the $u=T$ line. } \label{fig:parallel}
\end{figure}

In this section, we will use a formalism based upon the Riemann
tensor correlation function to calculate line broadening or
narrowing due to spacetime geometry fluctuations~\cite{TF06}.
Consider a source which emits signals at a mean frequency of
$\omega_0$ in its rest frame. The frequency detected by an observer
is subject to Doppler and gravitational redshifts, and the geometry
fluctuations will cause a fluctuation in the gravitational redshift.
Let \be \xi = \frac{\Delta \omega}{\omega_0} \ee
 be the fractional frequency
shift. Now consider two successive signals sent from the source to
the observer. The mean squared variation in $\xi$ between these two
signals due to geometry fluctuations is
\begin{equation}
  \delta\xi^2 = \langle(\Delta\xi)^2\rangle -
  \langle\Delta\xi\rangle^2 \,.
\end{equation}
In Ref.~\cite{TF06}, it is shown that this quantity may be expressed
in terms of the Riemann tensor correlation function
\begin{equation}\label{CorFunction}
 C_{\alpha\beta\mu\nu\,\gamma\delta\rho\sigma}(x,x') = \langle
 R_{\alpha\beta\mu\nu}(x)
 R_{\gamma\delta\rho\sigma}(x')\rangle -
 \langle R_{\alpha\beta\mu\nu}(x)\rangle
 \langle R_{\gamma\delta\rho\sigma}(x')\rangle \,.
\end{equation}
Specifically,
\begin{equation}
  \delta\xi^2 =
  \int da \int da' \,
  C_{\alpha\beta\mu\nu\,\gamma\delta\rho\sigma}(x,x')
  t^{\alpha}k^{\beta}t^{\mu}k^{\nu}t^{\gamma}k^{\delta}t^{\rho}k^{\sigma} \,.
                    \label{eq:delta_xi}
\end{equation}
Here $t^\mu$ is the four-velocity of both the source and detector,
which are assumed to be at rest with respect to one another, and
$k^\nu$ is the tangent to the worldlines of the signals. The
integrations in Eq.~(\ref{eq:delta_xi}) are taken over the region
bounded by the worldlines of the two signals and those of the source
and detector, as illustrated in Fig.~\ref{fig:parallel}.

In our present problem, the average geometry is that of Minkowski
spacetime, so that $ \langle R_{\alpha\beta\mu\nu}(x)\rangle = 0$,
and we have
 \ben
  \delta\xi^2 =\int da\int da'
 \langle R^{\mu}_{\alpha\nu\beta}R^{\rho}_{\gamma\sigma\lambda}\rangle
 t_{\mu}t_{\rho} t^{\nu}t^{\sigma}
 k^{\alpha}k^{\beta}k^{\gamma}k^{\lambda} \,.
 \een
Let $t^{\mu}=(1,0,0,0)$ and $k^{\mu}=(1,1,0,0)$, so that
 \ben
  \delta\xi^2 = \int da\int da'
 \langle R_{txtx}(x)R_{txtx}(x')\rangle \,.
 \een
The integrand in the above expression is given by
Eqs.~(\ref{eq:Rxtxt}) and (\ref{eq:Gx}) to be \ben
 \langle R_{txtx}R_{txtx}\rangle= \frac{1}{4}\,\partial_t^4G_{xxxx}=
\frac{n+1}{n+2}\,\partial_t^4 (D-2F_{xx}+H_{xxxx} =
\frac{n+1}{n+2}\,(\partial_t^2 - \partial_x^2)\, D   \,,
 \een
where in the last step we used $\partial_t^4H_{xxxx}=\partial_x^4D$
and $\partial_t^4F_{xx}=\partial_t^2\partial_x^2 D$. Define $u=t-x$
and $v=t+x$ and write the above expression as
 \ben
 \langle R_{txtx}R_{txtx}\rangle =
8 \,\frac{n+1}{n+2}\, \partial_u
\partial_{u'}\partial_v\partial_{v'}D.
 \een
Then we find
 \ben
 \delta \xi^2 &=&4\int da\int da'\partial_u \partial_{u'}\partial_v\partial_{v'}D
 \nonumber \\
&=& 8 \frac{n+1}{n+2}\,
\int_0^Tdu\int_0^Tdu'\int_u^{u+2r}dv\int_{u'}^{u'+2r}dv'\partial_v\partial_{v'}(\partial_u
\partial_{u'}D)\,.
 \een
Because $D$ depends on $v$ $ and $  $v'$ only through $\Delta
v=v-v'$, we find \ben
 \delta \xi^2 &=&
8 \frac{n+1}{n+2}\,\int_0^Tdu\int_0^Tdu' \Bigl[2(\partial_u
\partial_{u'}D)|_{\Delta v=\Delta u} \nonumber \\
&-& (\partial_u
\partial_{u'}D)|_{\Delta v=\Delta u+2r}- (\partial_u
\partial_{u'}D)|_{\Delta v=\Delta u-2r}\Bigr]\,.
\label{eq:xi2}
 \een

\subsection{ Four Dimensional Case}

Here $n=0$ and \ben D=\sum_{n=-\infty}^\infty{'}
\frac{1}{4\pi^2(\Delta x^2+n^2 L^2-\Delta
t^2)}=\frac{1}{2\pi^2}\sum_{n=1}^\infty \frac{1}{-\Delta v\Delta u+
n^2L^2}\,.\een

Use the relation \be\int_0^Tdu\int_0^Tdu'=2\int_0^Td\Delta u
(T-\Delta u)\,,\ee to write Eq.~(\ref{eq:xi2}) as \be \delta \xi^2 =
T_1+T_2+T_3 \,,\ee where
 \ben
T_1&=&16\int_0^Td\Delta u (T-\Delta u)(\partial_u
\partial_{u'}D)|_{\Delta v=\Delta u}\nonumber\\
&=&-\frac{16}{\pi^2}\sum_{n=1}^\infty \int_0^Td\Delta u (T-\Delta u)
\frac{\Delta u^2}{(-\Delta u^2+ n^2L^2)^3}\,, \een

 \ben
T_2&=&-8\int_0^Td\Delta u (T-\Delta u)(\partial_u
\partial_{u'}D)|_{\Delta v=\Delta u+2r}\nonumber\\
&=&-\frac{8}{\pi^2}\sum_{n=1}^\infty \int_0^Td\Delta u (T-\Delta
u)\frac{(\Delta u+2r)^2}{[-\Delta u(\Delta u+2r)+ n^2L^2)]^3}\,,
\een and
 \ben
T_3&=&-8\int_0^Td\Delta u (T-\Delta u)(\partial_u
\partial_{u'}D)|_{\Delta v=\Delta u-2r}\nonumber\\
&=&-\frac{8}{\pi^2}\sum_{n=1}^\infty \int_0^Td\Delta u (T-\Delta
u)\frac{(\Delta u-2r)^2}{[-\Delta u(\Delta u-2r)+ n^2L^2)]^3}\,.
\een

In the limits $r \gg T\gg L$, we obtain \be T_1\approx \frac{2}{3
L^2}\;,\ee
 \be T_2\approx \frac{T r}{2\pi^2L^4}\sum_{n=1}^\infty \frac{1}{n^4}-\frac{2}{3 L^2}\;,\ee
  \be T_3\approx -\frac{T r}{2\pi^2L^4}\sum_{n=1}^\infty \frac{1}{n^4}-\frac{2}{3 L^2}\;.\ee
Thus we find \be \delta \xi^2 \approx-\frac{2\,\ell_P^2}{3L^2}\,.\ee
The fact that $\delta \xi^2 < 0$ seems to imply a small narrowing of
spectral lines. Unless $L$ is very small,
 the natural line width of a spectral line is likely to be much larger in magnitude
than this effect. This narrowing is analogous to the negative shifts
in mean squared velocity of a charged or polarizable particle near a
boundary found in Refs.~\cite{WKF01,YF04}. Unlike $\Delta t$, this
effect does not grow with increasing path length.

\subsection{Five Dimensional Case}

Now we turn to the five dimensional model with one compactified
space dimension, which was studied in Sect.~\ref{sec:5DKK}. Here
\ben D=\frac{1}{4\pi^2}\, {\rm Re} \bigg[\sum_{n=1}^\infty
\frac{1}{(-\Delta v\Delta u+ n^2L^2)^{3/2}}\bigg]\,. \een As in the
four dimensional case, we may write
  \be
\delta \xi^2= \frac{4}{3}\,(T_1+T_2+T_3)\,, \ee where \be
T_1=16\int_0^Td\Delta u (T-\Delta u)(\partial_u
\partial_{u'}D)|_{\Delta v=\Delta u}\,, \ee

 \be
T_2=-8\int_0^Td\Delta u (T-\Delta u)(\partial_u
\partial_{u'}D)|_{\Delta v=\Delta u+2r}\,, \ee and
 \be
T_3=-8\int_0^Td\Delta u (T-\Delta u)(\partial_u
\partial_{u'}D)|_{\Delta v=\Delta u-2r}\,. \ee

In the limits $T\gg L$ and $r\gg T \& L$, we now obtain \be
T_1\approx \frac{2}{\pi^2}\sum_{n=1}^\infty
\frac{1}{n^3L^3}=\frac{2\xi (3)}{\pi^2 L^3}\;,\ee
 \be T_2\approx
\frac{2}{\pi^2}\sum_{n=1}^\infty\bigg[-\frac{3Tr}{n^5L^5}-\frac{1}{n^3L^3}\bigg]\;,\ee
  \be T_3\approx
\frac{2}{\pi^2}\sum_{n=1}^\infty\bigg[\frac{3Tr}{n^5L^5}-\frac{1}{n^3L^3}\bigg]\;.\ee
Thus we find in this case \be \delta \xi^2  \approx
-\frac{8\zeta(3)}{3\pi^2 L^3}\,, \ee where $\zeta(3) \approx 1.20$
is a Riemann zeta-function. We may use Eq.~(\ref{eq:Newton}) to
write this as \be \delta \xi^2    \approx -\frac{256 \zeta(3)\,
\ell_P^2}{3 \pi \, L^2}\,. \ee Again, we find a negative value for
$\delta \xi^2$ which does not increase as the flight path length
increases, and whose magnitude is determined by the ratio
$\ell_P/L$.

Let us comment on the relation between the result of
Sect.~\ref{sec:5DKK} that $\Delta t$ grows with increasing path
length, and the present result that $\delta \xi^2$ does not. If
the crests of a plane wave could be treated as truly uncorrelated
pulses, then we would expect the growth of $\Delta t$ to lead to
increasing line broadening with increasing $r$. When
Eq.~(\ref{eq:Leff}) is satisfied, the correlations between the
pulses becomes weak, but is never completely absent. Apparently
the remaining correlation is sufficient to prevent $\Delta \omega$
from increasing with increasing $r$.


\section{ Discussion and conclusions}
\label{sec:diss}


We have treated the effects of compact extra flat spacetime
dimensions on quantum lightcone fluctuations. One measure of 
lightcone fluctuations is variations in the flight times of pulses.
We found in a five dimensional model that this variation will grow
as the logarithm of the flight distance. In principle this is an
observable effect, but it is too small to be a realistic test of
extra dimensions. We also examined the correlations of successive
pulses. This correlation weakens as the pulse separation increases,
but is always nonzero. This nonzero correlation is presumably
responsible for the result that effect of geometry fluctuations on
the width of spectral lines does not increase with increasing flight
distance. In fact, we find a weak line narrowing effect in which the
spacetime geometry fluctuations slightly reduce the natural line
width. The fractional line narrowing effect is \be \frac{\Delta
\omega}{\omega} = \sqrt{ |\langle\delta \xi^2\rangle|} = C\,
\frac{\ell_P}{L}\,, \ee
 where $C$ is a constant of order unity.
The analysis in this paper assumes gravitons on a fixed background
spacetime and is presumable only valid when the compactification
scale is well above the Planck scale, $L \gg \ell_P$. Nonetheless,
this suggests a possible observational signature of the existence of
extra dimensions: a small, systematic narrowing of all spectral line
from what would otherwise be expected.

\begin{acknowledgments}
This work was supported in part by the National Natural Science
Foundation of China under Grants No. 10375023, 10775050, the SRFDP
under Grant No. 20070542002, the National Science Foundation under
 Grant PHY-0555754 and by Conselho Nacional de Desenvolvimento
Cientifico e Tecnologico do Brasil (CNPq).  LHF would like to
thank the Institute of Physics at Academia Sinica in Taipei,
National Dong Hwa University in Hualien, Taiwan  and the Kavali Institute
for Theoretical Physics China for hospitality
while this manuscript was completed. HY thanks the Kavali Institute
for Theoretical Physics China for hospitality in the final stages of this work.
\end{acknowledgments}

\appendix


\section{Graviton two-point functions}
\label{sec:2pt}


Here we evaluate the functions $D^n(x.x')$, $F^n_{ij}(x,x')$ and
$H^n_{ijkl}(x,x')$ defined in Eqs.~(\ref{eq:Ffunc}),
(\ref{eq:Dfunc}) and (\ref{eq:Hfunc}), respectively.
 Once these functions are given, the graviton
two point functions are easy to obtain. Define \be R=|{\bf x}-{\bf
x}'|, \quad \Delta t=t-t',\quad k=|{\bf k}|=\omega \,, \ee and
assume $n$  extra dimensions, then \ben D^{n}(x,x')&=&
{Re\over{(2\pi)^{3+n}}}\int\, {d^{(3+n)}{\bf k}\over{2
\omega}}e^{i{\bf k}
\cdot({\bf x}-{\bf x'})}e^{-i\omega(t-t')} \,\nonumber\\
&=&{Re\over{2(2\pi)^{3+n}}}\int_0^{\infty}\,k^{n+1}e^{-ik\Delta
t}\,dk \int_0^{\pi}\, d \theta_1\,\sin^{1+n}\theta_1 e^{i k R
\cos\theta_1} \nonumber\\&&\quad \times \int_0^{\pi}\, d
\theta_2\,\sin^{n}\theta_2... \int_0^{\pi}\, d
\theta_{n+1}\,\sin\theta_{n+1}
\int_0^{2\pi}\,d\theta_{n+2}\, \nonumber\\
&=&{a_n\,Re\over{2(2\pi)^{3+n}}}\,\int_0^{\infty}\,k^{n+1}e^{-ik\Delta
t}\,dk
\int_{-1}^{1}\,e^{ikRx}(1-x^2)^{n/2}\,dx\,\nonumber\\
&=&{a_n\,Re\over{(2\pi)^{3+n}}}\,\int_0^{\infty}\,k^{n+1}e^{-ik\Delta
t}\,dk
\int_{0}^{1}\,(1-x^2)^{n/2}\cos(kRx)\,dx\,\nonumber\\
&=&{a_n\sqrt{\pi}2^{{n+1\over
2}}\Gamma({n\over2}+1)\,Re\over{2(2\pi)^{3+n}}}\, {1\over
R^{n+1\over 2}}\int_0^{\infty}\,k^{n+1\over2}J_{n+1\over2}(kR)
e^{-ik\Delta t}\,dk\nonumber\\
&=&{a_n\sqrt{\pi}2^{{n+1\over
2}}\Gamma({n\over2}+1)\,Re\over{2(2\pi)^{3+n}}}\, {1\over
R^{n+1\over 2}}\lim_{\alpha\rightarrow 0^++i\Delta t}
\int_0^{\infty}\,k^{n+1\over2}J_{n+1\over2}(kR)e^{-\alpha k}\,dk\nonumber\\
&=&{a_n 2^n \Gamma({n\over2} +1)^2\over (2\pi)^{3+n}} {1\over
(R^2-\Delta t^2)^{n/2+1}} ={\Gamma({n\over2}+1)\over 4\pi^{
n+4\over2}}
{1\over (R^2-\Delta t^2)^{n/2+1}}\,.  \nonumber\\
\label{eq:Dn} \een Here we have defined \be a_n=\int_0^{\pi}\, d
\theta_2\,\sin^{n}\theta_2... \int_0^{\pi}\, d
\theta_{n+1}\,\sin\theta_{n+1} \int_0^{2\pi}\,d\theta_{n+2}={2\pi^{
{n\over2}+1}\over\Gamma({n\over2}+1)}\,, \ee and used \be
\int_{0}^{1}\,\cos(kRx)(1-x^2)^{n/2}\,dx= {\sqrt{\pi}\over
2}\Gamma({n\over2}+1)\biggl({2\over kR}\biggr)^{n+1\over
2}J_{n+1\over2}(kR)\,, \ee and \be \int_0^{\infty}e^{-\alpha
x}J_{\nu}(\beta x)x^{\nu}\,dx= {(2\beta)^{\nu}\Gamma(\nu+1/2)
\over\sqrt{\pi}(\alpha^2+\beta^2)^{\nu+{1\over2}}}\,, \quad\quad
{\rm Re}\, \nu> -1/2\,. \ee When  $n$ is odd,  $D^n(x,x')$ should be
taken to be zero when $R^2<\Delta t^2$.

Let us now turn our attention to the calculation of $F_{ij}$ and
$H_{jikl}$. We find \ben
F_{ij}^n(x,x')&=&{Re\over{(2\pi)^{3+n}}}\int\, d^{3+n}{\bf k}{k_ik_j
\over{2 \omega^3}}e^{i{\bf k} \cdot({\bf x}-{\bf
x'})}e^{-i\omega(t-t')}
\nonumber\\
&=&{Re\over{2(2\pi)^{3+n}}}\partial_i\partial_j'\int_0^{\infty}\,k^{n-1}
e^{-ik\Delta t}\,dk \int_0^{\pi}\, d \theta_1\,\sin^{1+n}\theta_1
e^{i k R \cos\theta_1} \nonumber\\&&\quad \times \int_0^{\pi}\, d
\theta_2\,\sin^{n}\theta_2... \int_0^{\pi}\, d
\theta_{n+1}\,\sin\theta_{n+1}
\int_0^{2\pi}\,d\theta_{n+2}\, \nonumber\\
&=&{a_n\,Re\over{(2\pi)^{3+n}}}\,\partial_i\partial_j'\int_0^{\infty}\,k^{n-1}
e^{-ik\Delta t}\,dk
\int_{0}^{1}\,(1-x^2)^{n/2}\cos(kRx)\,dx\,\nonumber\\
&=&{a_n\sqrt{\pi}2^{{n+1\over 2}}\Gamma({n\over2}+1)\,
Re\over{2(2\pi)^{3+n}}}\,\partial_i\partial_j'\left( {1\over
R^{n+1\over 2}}\int_0^{\infty}\,k^{n-3\over2}J_{n+1\over2}(kR)
e^{-ik\Delta t}\,dk\right)\nonumber\\
&=&{a_n\sqrt{\pi}2^{{n+1\over
2}}\Gamma({n\over2}+1)\over{2(2\pi)^{3+n}}}\,
\partial_i\partial_j'\left(
{Re\over R^{n+1\over
2}}\int_0^{\infty}\,k^{n-3\over2}J_{n+1\over2}(kR)
e^{-ik\Delta t}\,dk\right)\nonumber\\
&=&{Re\over 2(2\pi)^{3+n\over2}}\,\partial_i\partial_j'\left(
{1\over R^{n+1\over
2}}\int_0^{\infty}\,k^{n-3\over2}J_{n+1\over2}(kR)
e^{-ik\Delta t}\,dk\right)\nonumber\\
&=&{Re\over 2(2\pi)^{3+n\over2}}\,\partial_i\partial_j'\biggl(
{n-1\over R^2}{1\over R^{n-1\over 2}}\int_0^{\infty}\,k^{n-5\over2}
J_{n-1\over2}(kR)e^{-ik\Delta t}\,dk \nonumber\\&&\quad-{1\over
R^{n+1\over 2}}\int_0^{\infty}\,k^{n-3\over2}
J_{n-3\over2}(kR)e^{-ik\Delta t}\,dk\,\biggr)\,, \label{eq:Fn} \een
where we have utilized a recursive formula for Bessel functions \be
zJ_{\nu-1}(z)+zJ_{\nu+1}(z)=2\nu J_{\nu}(z)\,. \ee Similarly, one
finds that \ben H_{ijkl}^n(x,x')&&={Re\over{(2\pi)^{3+n}}}\int\,
d^{3+n}{\bf k}{k_i k_j k_k k_l
\over{2 \omega^5}}e^{i{\bf k} \cdot({\bf x}-{\bf x'})}e^{-i\omega(t-t')} \nonumber\\
&&={Re\over
2(2\pi)^{3+n\over2}}\,\partial_i\partial_j'\partial_k\partial_l\biggl(
{n-1\over R^2}{1\over R^{n-1\over
2}}\int_0^{\infty}\,k^{n-9\over2}J_{n-1\over2}(kR)e^{-ik\Delta
t}\,dk \nonumber\\&&\quad-{1\over R^2}{1\over R^{n+1\over
2}}\int_0^{\infty}\,k^{n-7\over2}J_{n-3\over2}(kR) e^{-ik\Delta
t}\,dk\,\biggr)\,. \label{eq:Hn} \een To proceed further with the
calculation, we need to deal with the cases when $n$ is odd or even
separately.


\subsection{The case of odd n}


 Assume $n=2m+1$ and define
\be S(m)={Re\over
R^{m+1}}\,\int_0^{\infty}\,k^{m-1}J_{m+1}(kR)e^{-ik\Delta
t}\,dk\,,\quad m\geq 0\,, \ee \ben
T(m-1)\,&=&{Re\over R^{m+1}}\,\int_0^{\infty}\,k^{m-1}J_{m-1}(kR)e^{-ik\Delta t}\,dk\nonumber\\
&=&{Re\over R^{m+1}}\,\lim_{\alpha\rightarrow 0^++i\Delta
t}\int_0^{\infty}\,k^{m-1}J_{m-1}(kR)
e^{-alpha k}\,dk\nonumber\\
&=&{2^{m-1}\Gamma(m-1/2)\over\sqrt{\pi}}{\sqrt{R^2-\Delta t^2}\over R^2(R^2-\Delta t^2)^m}\,,\nonumber\\
&=&{(2m-1)!!\over (2m-1)} {\sqrt{R^2-\Delta t^2}\over R^2(R^2-\Delta
t^2)^m}\,,\quad m\geq 1\,, \een where we have appealed to integral
(6.623.1) in Ref.~\cite{GR1}.The above result holds for $R^2>\Delta
t^2$, and $T(m-1)$ is zero when $R^2<\Delta t^2$. Then it follows
from Eq~(\ref{eq:Fn}) that \be F^{2m+1}_{ij}={1\over
2(2\pi)^{m+2}}\,\partial_i\partial_j'\biggl(S(m)\biggr)\,,
                                               \label{eq:F}
\ee and \be S(m)={2m\over R^2}S(m-1)-T(m-1) \label{eq:Recursive1}
\ee

Using the recursive relation Eq~(\ref{eq:Recursive1}), we can show
that \ben
S(m)\,&=&{(2m)!!\over R^{2m}}S(0)-\sum_{k=1}^m{(2m)!!\over (2k)!!}{T(k-1)\over R^{2m-2k}}\nonumber\\
&=&{(2m)!!\over R^{2m}}S(0)\left[1-\sum_{k=1}^m{(2k-1)!!\over
(2k)!!(2k-1)}
{R^{2k}\over (R^2-\Delta t^2)^k}\right]\nonumber\\
&=&\,-{(2m)!!\over R^{2m}}S(0)\,\sum_{k=0}^m{(2k+1)!!\over
(2k)!!(2k+1)(2k-1)} {R^{2k}\over (R^2-\Delta t^2)^k}\,. \een Here
\begin{eqnarray}
S(0)&=& {Re\over R}\lim_{\alpha\rightarrow 0^++i\Delta
t}\int_0^{\infty}\,{J_1(kR)\over k} e^{-\alpha k}dk
    = {1\over R}\int_0^{\infty}\,{J_1(kR)\cos(k\Delta t)\over k}\,dk
\,\nonumber\\
&=& \left\{\begin{array}{ll}
             {1\over R}\cos\biggl(\arcsin(\Delta t/R)\biggr)={\sqrt{R^2-\Delta t^2}\over R^2}
                 &\mbox{ for}\, R^2>\Delta t^2\,,\\
             0 &\mbox{ for}\, R^2<\Delta t^2\,.
    \end{array}\right.
\label{eq:S0}
\end{eqnarray}

If we define \be Q(m)={Re\over
R^{m+1}}\,\int_0^{\infty}\,k^{m-3}J_{m+1}(kR) e^{-ik\Delta
t}\,dk\,,\quad m\geq 0\,, \ee then it is easy to see that
 \be
 \label{H}
H_{ijkl}^{2m+1}={1\over 2(2\pi)^{m+2}}\,\partial_i\partial_j'
\partial_k\partial_l'\biggl(Q(m)\biggr)\,,
\label{eq:H} \ee and \be Q(m)={2m\over R^2}Q(m-1)-{1\over
R^2}S(m-2)\,. \label{eq:Recursive2} \ee The above equation applies
for $m\geq 2$. To use it to get a general expression,
 we need $Q(0)$, which can be calculated in the case of $R^2<\Delta
 t^2$ as follows
\ben Q(0)&=&{1\over R}\int_0^{\infty}\,
{1\over k^3}J_1(kR)\cos(k\Delta t)\,dk\,\nonumber\\
&=& \lim_{\beta\rightarrow 0}\, {1\over R}\int_0^{\infty}\,
{k^{-1}\over (k^2+\beta^2)}J_1(kR)\cos(k\Delta t)\,dk\,\nonumber\\
&=&\lim_{\beta\rightarrow 0}\, {1\over R}{e^{-\beta\Delta
t}I_1(\beta R)\over \beta^2}
={1\over 2\beta}-{1\over 2\Delta t}\,\,\nonumber\\
\label{eq:Q0} \een
 This leads to a vanishing $H_{ijkl}$. However, in the case of $R^2>\Delta
 t^2$, the calculation becomes a little complicated. First, let us
 write $Q(0)$ as
 \ben
 \label{Q0}
Q(0)&=&-{1\over 2R} \int_0^\infty\,\biggl({1\over
k^2}\biggr)^{'}J_1(kR)
\cos(k\Delta t)\,dk \nonumber\\
&=&\lim_{k\rightarrow 0}-{1\over 4k}\cos(k\Delta t)+P_1+P_2-{1\over
2}Q(0)\;,
 \een
where we have used the fact that
 \ben
 J^{'}_1(x)&=& J_0(x)-{1\over x} J_1(x)\;,\nonumber\\
 J_1(x)&\sim &{x\over 2},\;\;\; {\text as}\;\;\; x \rightarrow 0
 \;,
 \een
and defined
 \be
P_1=-{\Delta t\over 2R}\,\int_0^{\infty}\,{1\over
k^2}\,J_1(kR)\sin(k\Delta t)\,dk \;,
 \ee
and
 \be
 P_2= {1\over
2}\,\int_0^{\infty}\,{\cos(k\Delta t)J_0(kR)\over k^2}\,dk\;.
 \ee
 Note that the first term in Eq.~(\ref{Q0}) can be dropped although it
 is formally divergent. The reason is that it is only dependent on
 $\Delta t$ and $H_{ijkl}$ involves spatial differentiation. So, it
 follows that
 \be
 Q(0)={2\over 3}(P_1+P_2)\;.
 \ee
 Our next task is then to evaluate $P_1$ and $P_2$, which can be
 done as follows
  \ben
P_1&=&{\Delta t\over 2R}\,\int_0^{\infty}\,\biggl({1\over
k}\biggr)^{'}J_1(kR)\sin(k\Delta t)\,dk \nonumber\\
&=&-{\Delta t^2\over 2R}\,\int_0^{\infty}\,J_1(kR)\cos(k\Delta t){dk
\over k}-{\Delta t\over 2}\,\int_0^{\infty}\,J_0(kR)\sin(k\Delta
t){dk \over k}-P_1\;.
 \een
Substitution of integrals (6.693.1) and (6.693.2) in Ref.~\cite{GR1}
into the above equation yields
 \be
 P_1=-{\Delta t ^2\over 4R^2}\sqrt{R^2-\Delta t^2}-{\Delta t \over
 4}\arcsin(\Delta t/R)\;.
\ee
 As for $P_2$, the calculation goes
 \ben
 P_2&=&-{1\over 2}\,\int_0^{\infty}\,\biggl({1\over
k}\biggr)^{'} \cos(k\Delta t) J_0(kR)\,dk\,\nonumber\\
&=&-{\Delta t \over 2}\,\int_0^{\infty}\,J_0(kR)\sin(k\Delta t){dk
\over k}-{R\over 2}\,\int_0^{\infty}\,J_1(kR)\cos(k\Delta t){dk
\over k}\nonumber\\
&=&-{\Delta t \over 2}\,\arcsin(\Delta t/R)-{\sqrt{R^2-{\Delta
t^2}}\over 2}\;
 \een
 where we have also discarded a formally divergent term dependent
 only on $\Delta t$ since it does not contribute to $H_{ijkl}$. A
 combination of the above derived results finally leads to
 \begin{equation}
 \label{Q0F}
Q(0) = -{ \sqrt{R^2-\Delta t^2}(\Delta t^2+2R^2)\over 6R^2}-{\Delta
t\over 2}\arcsin(\Delta t/R)\;.
\end{equation}

We next  need  $Q(1)$, which can be calculated, using integral
(6.693.5) in Ref.~\cite{GR1},
  as follows
 \ben
 \label{Q1}
Q(1)&=&{1\over R^2}\int_0^{\infty}\, {J_2(Rk)\cos(\Delta t k)\over k^2}\,dk\nonumber\\
&=&
  \left\{\begin{array}{ll}
       {1\over R^2}\left[ {R\over 4}\cos(\arcsin(\Delta t/R))+{R\over 12}\cos(3\arcsin(\Delta t/R))\right]
      &\mbox{ for}\, R^2>\Delta t^2\,,\\
      0 &\mbox{ for}\, R^2<\Delta t^2\,.
   \end{array}\right.
\nonumber\\
&=&
   \left\{\begin{array}{ll}
           \biggl({1\over3}-{\Delta t^2\over 3R^2}\biggr){\sqrt{R^2-\Delta t^2}\over R^2}
                =\biggl({1\over3}-{\Delta t^2\over 3R^2}\biggr)S(0) &\mbox{ for}\, R^2>\Delta t^2\,,\\
          0 &\mbox{ for}\, R^2<\Delta t^2\,.
    \end{array}\right. \nonumber\\
\een In the above calculation, we have made use of the following
trigonometric relations \be
 \cos(3x)=4\cos^3(x)-3\cos(x), \quad\quad \cos(\arcsin x)=\sqrt{1-x^2}.\nonumber\\
\ee
 Therefore, for $m\geq 2$, one finds, using the recursive relation
Eq~(\ref{eq:Recursive2}), \ben
Q(m)&=&{(2m)!!\over 2R^{2m-2}}Q(1)-{1\over R^2}\sum_{k=2}^m{(2m)!!\over (2k)!!}{S(k-2)\over R^{2m-2k}}\nonumber\\
&=&{(2m)!!\over R^{2m-2}}\Biggl[{1\over2}Q(1)\nonumber\\
&& +\,\sum_{k=2}^m\sum_{j=0}^{k-2} {(2j+1)!!\over
2k(2k-2)(2j)!!(2j+1)(2j-1)} {R^{2j}\over (R^2-\Delta
t^2)^j}S(0)\,\Biggr]\,.
\nonumber\\
\een This  expression can be simplified  if we note that \be
\sum_{k=j+2}^m\, {1\over k(k-1)}= \sum_{k=2}^m\, {1\over
k(k-1)}-\sum_{k=2}^{j+1}\, {1\over k(k-1)} ={{ m-j-1}\over
m(j+1)}\,, \ee and \ben
 &&\sum_{k=2}^m\sum_{j=0}^{k-2}
{(2j+1)!!\over 2k(2k-2)(2j)!!(2j+1)(2j-1)}
{R^{2j}\over (R^2-\Delta t^2)^j}S(0)\,\Biggr]\nonumber\\
&&\quad=\sum_{j=0}^{m-2}\sum_{k=j+2}^m {(2j+1)!!\over
2k(2k-2)(2j)!!(2j+1)(2j-1)}
{R^{2j}\over (R^2-\Delta t^2)^j}S(0)\,\Biggr]\nonumber\\
&&\quad=\sum_{j=0}^{m-2} {(m-j-1)(2j+1)!!\over
4m(j+1)(2j)!!(2j+1)(2j-1)}
{R^{2j}\over (R^2-\Delta t^2)^j}S(0)\,\Biggr]\nonumber\\
\een So, we have in this case \be D^{2m+1}=
  \left\{\begin{array}{ll}
            {(2m+1)!!\over 2(2\pi)^{m+2}}{1\over (R^2-\Delta t^2)^{m+{3\over2}}}\,,
                      &\mbox{ for}\, R^2>\Delta t^2\,,\\
             0 &\mbox{ for}\, R^2<\Delta t^2\,,
   \end{array}\right.
\ee \be F^{2m+1}_{ij}=-{1\over
2(2\pi)^{m+2}}\,\partial_i\partial_j'\left( {(2m)!!\over
R^{2m}}S(0)\,\sum_{k=0}^m{(2k+1)!!\over (2k)!!(2k+1)(2k-1)}
{R^{2k}\over (R^2-\Delta t^2)^k}\right)\,, \ee and \ben
H_{ijkl}^{2m+1}&=&{1\over
2(2\pi)^{m+2}}\,\partial_i\partial_j'\partial_k\partial_l'\biggl\{
 {(2m)!!\over R^{2m-2}}\biggl[{1\over2}Q(1)\nonumber\\
&&+\,\sum_{j=0}^{m-2} {(m-j-1)(2j+1)!!\over
4m(j+1)(2j)!!(2j+1)(2j-1)}
{R^{2j}\over (R^2-\Delta t^2)^j}S(0)\,\Biggr]\,,\nonumber\\
\een
 for $m\geq 2$, while for $m=0$ and $m=1$, $H_{ijkl}$ can be found
 by using Eq.~(\ref{H}), Eq.~(\ref{Q0F}) and Eq.~(\ref{Q1})


\subsection{The case of even n}


Let $n=2m$ with $m=1,2,3...$. The graviton two-point functions for
$m=0$ corresponding to the usual 4 dimensional spacetime have been
given previously \cite{YUF}. The analog of Eq.~(\ref{eq:F}) for this
case is \be F^{2m}_{ij}={1\over 2(2\pi)^{m+{3\over2}}}\,
\partial_i\partial_j'\biggl(S(m-{1\over2})\biggr) \,.
\ee Here \be S(m-1/2)={2m-1\over
R^2}S(m-{3\over2})-T(m-{3\over2})\,. \label{eq:Recursive3} \ee Using
this recursive relation, we can express $S(m-1/2)$ in terms of
S(1/2) which is
 calculated, by employing
\be
J_{n+{1\over2}}(z)=(-1)^n\,z^{n+{1\over2}}\sqrt{{2\over\pi}}{d^n\over(zdz)^n}\biggl(
{\sin z\over z}\,\biggr)\,, \ee
 to be
\ben S(1/2)&&={1\over
R^{3\over2}}\int_0^{\infty}\,k^{-1/2}J_{3\over2}(Rk)\cos(\Delta t
k)\,dk
\nonumber\\
&&=-\sqrt{{2\over\pi}}{1\over R^3}\int_0^{\infty}\,{d\over
dk}\biggl({\sin (Rk)\over k}
\biggr)\cos(\Delta t k)\,dk\nonumber\\
&&=-\sqrt{{2\over\pi}}\left({1\over R^3}{\sin (Rk)\cos(\Delta t
k)\over k} \Bigg |_0^{\infty}+ {\Delta t\over
R^3}\int_0^{\infty}\,{\sin (Rk)\sin(\Delta t k)\over k}\,dk
\,\right)\nonumber\\
&&=\sqrt{{2\over\pi}}\left({1\over R^2}-{\Delta t\over4
R^3}\ln\biggl({R+\Delta t\over R-\Delta t}\biggr)^2\,\right)\,. \een
It then follows that \ben &&S(m-1/2)\,={(2m-1)!!\over
R^{2m-2}}S(1/2)-\sum_{k=2}^m{(2m-1)!!\over (2k-1)!!}
{T(k-3/2)\over R^{2m-2k}}\nonumber\\
&&={(2m-1)!!\over R^{2m}}\sqrt{{2\over\pi}}\left[1-{\Delta t\over4
R}\ln\biggl({R-\Delta t\over R-\Delta t}\biggr)^2\, -{1\over
R^2}\sum_{k=2}^m{2^{k-2}\Gamma(k-1)\over (2k-1)!!} {R^{2k}\over
(R^2-\Delta t^2)^{k-1}}\right]\,.
\nonumber\\
\een Similarly, one has for $H_{ijkl}^{2m}$ \be
H^{2m}_{ijkl}={1\over
2(2\pi)^{m+{3\over2}}}\,\partial_i\partial_j'\partial_k\partial_l'
\biggl(Q(m-{1\over2})\biggr)\,, \ee and \be Q(m-1/2)={2m-1\over
R^2}Q(m-{3\over2})-{1\over R^2}S(m-{5\over2})\,.
\label{eq:Recursive4} \ee

Now the calculation becomes a little tricky. First, let us note that
$H_{ijkl}^0$ has already been given \cite{YUF} and the recursive
relation Eq~(\ref{eq:Recursive4}) can only be applied when $m\geq
3$. So, we need both $H_{ijkl}^2$ and $H_{ijkl}^4$ or $Q(1/2)$ and
$Q(3/2)$ as our basis to use the recursive relation for a general
expression.
 Because there is an infrared divergence in the $Q(1/2)$
 integral, so, as we did in the 4 dimensional case,
we will introduce a regulator $\beta$ in the denominator of the
integrand and then let $\beta$ approach 0 after the integration is
performed. Noting that \be J_{3\over2}(z)=\sqrt{2\over\pi
z}\biggl({\sin z\over z}-\cos z\biggr)\,, \ee we obtain \ben
Q(1/2)&=&{1\over R^{3\over2}}\,\int_0^{\infty}\,k^{-5\over2}
J_{3\over2}(kR)\cos(kt)\,dk\,
\nonumber\\
&&=\sqrt{2\over\pi }\biggl({1\over R^3} \int_0^{\infty}\,{dk\over
k^4} \sin kR\, \cos k\Delta t\,-{1\over R^2}
\int_0^{\infty}\,{dk\over k^3}
 \cos kR\, \cos k\Delta t\,\biggr)
\nonumber\\
&&=\sqrt{2\over\pi }\lim_{\beta\rightarrow 0}\biggl(-{1\over R^3}
{1\over 2\beta}{\partial\over\partial\beta}
 \int_0^{\infty}\,{\sin kR \cos k\Delta t\over k^2+\beta^2}\,dk
\,\nonumber\\\quad\quad &&\quad +{1\over R^2}{1\over
2\beta}{\partial\over\partial\beta} \int_0^{\infty}\, {k\cos kR\cos
k\Delta t\over k^2+\beta^2} \,dk \,\biggr)\,.
\nonumber\\
\een We next use \ben \int_0^{\infty}\,{\sin(ax)\cos(bx)\over
\beta^2+x^2}\,dx=&& {1\over4\beta}e^{-a\beta}
\{e^{b\beta}Ei[\beta(a-b)]+e^{-b\beta}Ei[\beta(a+b)]\}\nonumber\\
&&-{1\over4\beta}e^{a\beta}
\{e^{b\beta}Ei[-\beta(a+b)]+e^{-b\beta}Ei[-\beta(a-b)]\}\,, \een
\ben \int_0^{\infty}\,{x\cos(ax)\cos(bx)\over \beta^2+x^2}\,dx=&&
-{1\over4}e^{-a\beta}
\{e^{b\beta}Ei[\beta(a-b)]+e^{-b\beta}Ei[\beta(a+b)]\}\nonumber\\
&&-{1\over4}e^{a\beta}
\{e^{b\beta}Ei[-\beta(a+b)]+e^{-b\beta}Ei[-\beta(a-b)]\}\,, \een
where ${\rm Ei(x)}$ is the exponential-integral function, and the
fact that, when $x$ is small, \be {\rm Ei(x)}\approx \gamma
+\ln|x|+x+{1\over 4}x^2+{1\over 18}x^3+O(x^4)\,, \ee where $\gamma$
is the Euler constant. After expanding $Q(1/2)$ around $\beta=0$ to
the order of $\beta^2$, one finds \ben
Q(1/2)&=&\lim_{\beta\rightarrow0}\sqrt{2\over \pi}
\Biggl({5\over18}-{1\over3}\gamma-{1\over3}\ln(\beta)
-{1\over6}\ln(R^2-\Delta t^2)\nonumber\\
&& -{\Delta t^2\over6R^2}+{\Delta t\over8R}\biggl({\Delta t^2\over3
R^2}-1\biggl) \ln\biggl({R+\Delta t\over R-\Delta t}\biggr)^2
\Biggr)\,. \een Note, however, that what we need is $H_{ijkl}$ which
involves differentiation of $Q(1/2)$, therefore we can discard the
constant and divergent terms in $Q(1/2)$ as far as $H_{ijkl}$ is
concerned. To calculate $Q(3/2)$, let us recall that \be
Q(3/2)={3\over R^2}Q(1/2)-{1\over R^2}S(-1/2) \ee
 and note that $S(-1/2)$ is given by $\sqrt{2/\pi}$ times Eq~(A19) in Ref.~\cite{YUF}
Thus, we have \be Q(3/2)=\sqrt{2\over \pi}\left(-{1\over6R^2}
-{\Delta t^2\over2R^4}+{\Delta t\over8R^3}\biggl({\Delta t^2\over
R^2}-1\biggl) \ln\biggl({R+\Delta t\over R-\Delta
t}\biggr)^2\right)\,. \ee With $Q(3/2)$ at hand, it is easy to show
that for an arbitrary $m\geq 3$ \ben Q(m-1/2)&=&{(2m-1)!!\over
3R^{2m-4}}Q(3/2)-
{1\over R^2}\sum_{k=3}^m{(2m-1)!!\over (2k-1)!!}{S(k-5/2)\over R^{2m-2k}}\nonumber\\
&=&{(2m-1)!!\over R^{2m-4}}\biggl(\,{1\over 3}Q(3/2)-\sum_{k=3}^m\,
{1\over(2k-1)(2k-3)}S(1/2)
\nonumber\\
&& +{1\over R^4}\sqrt{2\over\pi}\sum_{k=3}^m\sum_{j=2}^{k-2}\,
{2^{j-2}\Gamma(j-1)\over (2j-1)!!(2k-1)(2k-3)}{R^{2j}\over
(R^2-\Delta t^2)^{j-1}}\,
\biggr)\nonumber\\
&=&{(2m-1)!!\over R^{2m-4}}\biggl(\,{1\over 3}Q(3/2)-\sum_{k=3}^m\,
{1\over(2k-1)(2k-3)}S(1/2)
\nonumber\\
&& +{1\over R^4}\sqrt{2\over\pi}\sum_{j=2}^{m-2}\,
{(m-j-1)2^{j-2}\Gamma(j-1)\over (2m-1)(2j+1)!!}{R^{2j}\over
(R^2-\Delta t^2)^{j-1}}\,
\biggr)\,.\nonumber\\
\een Here in the last step, we have made use of the following
results \ben \sum_{k=j+2}^m\, {1\over(2k-1)(2k-3)}&=&
\sum_{k=2}^m\, {1\over (2k-1)(2k-3)}-\sum_{k=2}^{j+1}\, {1\over (2k-1)(2k-3)}\nonumber\\
&=&{{ m-j-1}\over (2m-1)(2j+1)}\,, \een and \be
\sum_{k=3}^m\sum_{j=2}^{k-2}\,f(j)g(k)=\sum_{k=4}^m\sum_{j=2}^{k-2}\,f(j)g(k)
=\sum_{j=2}^{m-2}f(j)\sum_{k=2+j}^{m}\,g(k)\,. \ee Consequently, we
obtain \be D^{2m}={2^m m!\over (2\pi)^{m+2}}{1\over (R^2-\Delta
t^2)^{m+1}}\,, \ee \ben F^{2m}_{ij}&=&{1\over
(2\pi)^{m+2}}\,\partial_i\partial_j'\biggl\{
{(2m-1)!!\over R^{2m}}\biggl[1-{\Delta t\over4 R}\ln\biggl({R-\Delta t\over R-\Delta t}\biggr)^2\,\nonumber\\
&& -{1\over R^2}\sum_{k=2}^m{2^{k-2}\Gamma(k-1)\over (2k-1)!!}
{R^{2k}\over (R^2-\Delta t^2)^{k-1}}\biggr]\biggr\}\,, \een and \ben
H_{ijkl}^{2m}&=&{1\over (2\pi)^{m+2}}\,
\partial_i\partial_j'\partial_k\partial_l'\biggl\{{(2m-1)!!\over R^{2m-4}}\biggl[\,
{\Delta t\over24R^3}\biggl({\Delta t^2\over R^2}-1\biggl)
\ln\biggl({R+\Delta t\over R-\Delta t}\biggr)^2
-{1\over18R^2}\nonumber\\
&&\quad\quad -{\Delta t^2\over6R^4} -\sum_{k=3}^m\,{1\over
(2k-1)(2k-3)}\left({1\over R^2}- {\Delta t\over4
R^3}\ln\biggl({R+\Delta t\over R-\Delta t}\biggr)^2\,\right)
\nonumber\\
&&\quad \quad+{1\over R^4}\sum_{j=2}^{m-2}\,
{(m-j-1)2^{j-2}\Gamma(j-1)\over (2m-1)(2j+1)!!}{R^{2j}\over
(R^2-\Delta t^2)^{j-1}}\,
\biggr)\,.\nonumber\\
\een


\section{Derivation of Eq.~(\ref{eq:sigma_sigma})}
\label{sec:sigma_sigma}

Here we wish to give the details of the derivation of
Eq.~(\ref{eq:sigma_sigma}),
 the expression for
$ \langle \sigma_1\sigma'_1\rangle$ in the limit of large $r$. The
relevant graviton two-point function can be expressed as \ben
&&G_{xxxx}(t,x,0,0,0, t', x',0,0, nL)|_{t-t'=\rho-T}=
\nonumber\\
&&\quad{1\over 3\pi^2}
{n^4L^4-10n^2L^2\beta(\rho,T)^2-15\beta(\rho,T)^4\over
\beta(\rho,T)^3(\rho^2+n^2L^2)^2} + {4\over 3\pi^2} {3n^4L^4
+20n^2L^2\beta(\rho,T)^2+5\beta(\rho,T)^4\over \beta(\rho,T)(\rho^2+n^2L^2)^3}\nonumber\\
&&\quad-{4\over 3\pi^2} {18n^4L^4\beta(\rho,T)
+20n^2L^2\beta(\rho,T)^3\over(\rho^2+n^2L^2)^4}+{64\over 3\pi^2}
{n^4L^4\beta(\rho,T)^3\over (\rho^2+n^2L^2)^5} \,,\een where \ben
\beta(\rho,T)\equiv \sqrt{n^2L^2+2\rho T-T^2}\,.\een Utilizing the
following integration relation \be \int_a^b\,dx\int_a^b\,dx'
f(x-x')= \int_0^r\, (r-\rho)[f(\rho)+f(-\rho)]\,d\rho\,, \ee
 one finds that
\ben \langle \sigma_1\sigma'_1\rangle&=&-{r^2\over 36\pi^2}
\sum_{n=1}^\infty \bigg[-{2(4n^2L^2-T^2)\sqrt{n^2L^2-T^2}\over n^4
L^4}+ {h(r,T)+h(-r,T)\over (n^2 L^2+r^2)^3}\bigg]\;,\een where \ben
h(r,T)\equiv \beta(r,T)[4n^4L^4+n^2L^2(6r-T)(2r+T)+3r^2(2r+T)^2] \,,
\een

A few things are to be noticed here: (1) We need to drop the terms
when the square root is imaginary. (2) It can be shown that the
above expression for $\langle \sigma_1\sigma'_1\rangle$ reduces to
$\langle \sigma_1^2\rangle $
 when $T=0$, as it should.
(3) The asymptotic behavior of the summand when $n\rightarrow
\infty$,  is
 $\sim{ 1\over n^3}$  hence the summation converges.

To proceed, let us now assume that $r\gg T, L$, then
 \ben
 \label{eq:A}
 \langle
\sigma_1\sigma'_1\rangle&\approx&-{r^2\over 36\pi^2}
\sum_{n=1}^\infty \bigg[-{2(4n^2L^2-T^2)\sqrt{n^2L^2-T^2}\over n^4
L^4}\nonumber\\
 &+&  {\sqrt{n^2L^2-2rT}\over (n^2
L^2+r^2)^3} [4n^4L^4+12r^2 n^2L^2+12r^4]\nonumber\\
&+&{\sqrt{n^2L^2+2rT}\over (n^2 L^2+r^2)^3}[4n^4L^4+12r^2
n^2L^2+12r^4] \bigg]\,, \een
  and
  \be p\equiv
\sqrt{{2rT\over L^2}} \gg 1 \label{eq:p} \ee
 is a huge number. Thus for the second term in Eq. (\ref{eq:A}),
the sum should only start from $n = p$. We can now split the
summation into two parts, i.e. terms with $n\leq p$ and those with
$n>p$. Using the asymptotic form of the summand for the part with
$n>p$ and defining $m=[T/L]$, where $[\;\;]$ denotes the integer
part, one has
 \ben
 \langle \sigma_1\sigma'_1\rangle &\approx&
{r^2\over 36\pi^2L }\biggl(
\sum^p_{n=m}{2(4n^2-m^2)\sqrt{n^2-m^2}\over n^4 }-
\sum^p_{n=1}{4\sqrt{n^2+p^2}\over (n^2+r^2/L^2)^3}[n^4+3n^2r^2/L^2
+3r^4/L^4]\nonumber\\
 &+&\sum^{\infty}_p{2T^2\over 3\pi^2 n^3L^3} \biggr)\,. \een
Hence, it follows that
 \ben
\langle \sigma_1\sigma'_1\rangle &<&{r^2\over 36\pi^2L }\biggl(
\int_m^p{2(4n^2-m^2)\sqrt{n^2-m^2}\over n^3 }+
\sum^p_{n=1}{4\sqrt{n^2+p^2}\over (n^2+r^2/L^2)^3}[n^4+3n^2r^2/L^2
+3r^4/L^4]\nonumber\\ &&+\sum^{\infty}_p{2T^2\over 3\pi^2 n^3L^3}
\biggr)\,.\een Let us now evaluate the above expression term by
term. One has, keeping in mind that $p\gg 1$, that
 \ben
\int_m^p{2(4n^2-m^2)\sqrt{n^2-m^2}\over n^3 }\approx 8 \ln {2p\over
m}\approx 4\ln {2r\over T}\,,\een

\ben &&\sum^p_{n=1}{4\sqrt{n^2+p^2}\over
(n^2+r^2/L^2)^3}[n^4+3n^2r^2/L^2 +3r^4/L^4]\nonumber\\
&&\qquad\approx \int_{1/p}^1 dx {\sqrt{1+x^2}\over
(x^2+r/2T)^3}(4x^4+6x^2r/T+3r^2/T^2)\nonumber\\
&&\qquad\approx 12[\sqrt{2}+ {\rm Coth}^{-1}(\sqrt{2})]{T\over
r}\,,\een
  and
 \be\sum^{\infty}_p{1\over n^3}=-{1\over 2}\Psi(2,p)\sim {1\over 2}{1\over p^2}
={1\over 4}{L\over r}{L\over T}\,. \ee Here we have used
Eq.~(\ref{eq:p}) and the asymptotic expansion for $\Psi(2,x)$ \be
\Psi(2,x)\approx -{1\over x^2}-{1\over x^3}-{1\over
2x^4}+O(1/x^6)\,, \ee
 where function $\Psi(n,x)$ is defined as
\be \Psi(n,x)={d^n \psi(x)\over dx^n}\,, \quad\quad \psi(x)={d\over
dx}\ln\Gamma \ee
  Keeping only the dominant terms, we obtain Eq.~(\ref{eq:sigma_sigma}).

\end{document}